\documentstyle[twoside,fleqn,espcrc2,epsf]{article}

\newcommand{\AmS}{{\protect\the\textfont2
  A\kern-.1667em\lower.5ex\hbox{M}\kern-.125emS}}
\newcommand{\pslash}{\not{\hspace{-0.08cm}p}}

\title{Lattice Renormalization of Quark Operators\thanks{Talk 
          given by H. Oelrich at Lat97, Edinburgh, U.K.} }

\author{M. G\"ockeler\address{Institut f\"ur Theoretische Physik,
        Universit\"at Regensburg, D-93040 Regensburg, Germany}%
        , R. Horsley\address{Institut f\"ur Physik, Humboldt-Universit\"at
        zu Berlin, D-10115 Berlin, Germany}%
        , H. Oelrich\address{DESY-IfH Zeuthen, D-15735 Zeuthen,
        Germany}%
        , H. Perlt\address{Institut f\"ur Theoretische Physik,
        Universit\"at Leipzig, D-04109 Leipzig, Germany}%
        , P. Rakow$^{\rm c}$%
        , G. Schierholz$^{\rm c,}$
           \hspace{-0.2cm}
        \address{Deutsches Elektronen-Synchrotron DESY,
        D-22603 Hamburg, Germany}%
        and A. Schiller$^{\rm d}$ 
        }
\begin{document}

\begin{abstract}
We have technically improved the non-perturbative renormalization 
method, proposed by Martinelli et al., by using quark 
momentum sources and sinks. Composite two-fermion operators up 
to three derivatives have been measured for Wilson fermions and
Sheikholeslami-Wohlert improved fermions in the quenched 
approximation. The calculations are performed in the Landau gauge
on $16^332$ lattices at $\beta = 6.0$ for $3$ $\kappa$ values in each case.
The improved sources greatly decrease the statistical noise.
We extract and discuss here renormalization factors for local operators
and moments of the structure functions for Wilson
fermions.
\end{abstract}

\maketitle

\section{INTRODUCTION}

In this paper we want to discuss some general and technical aspects
of calculating non-perturbatively renormalization factors of bilinear
quark operators by imposing
renormalization conditions on off-shell quark Green's functions 
on the lattice \cite{1}.
  
This method offers the possibility of computing 
non-perturbative contributions and all orders of $QCD$ perturbation theory
in a simple way in contrast to the enormous efforts to get only 
lower orders with perturbative methods. 
The disadvantages of this numerical renormalization are the need of
computer power and at present the systematic uncertainties 
due to a quenched approximation (that excludes
fermion loops), discretization errors \cite{2} and gauge fixing.

We look here at operators \cite{3} 
${\cal O} \sim \bar{q} O q$ 
with quark fields $q$ and $O$ constructed out of
Dirac operators $\gamma_\mu$ and covariant lattice derivatives
$D_\nu$.
\vspace*{-19cm} \\ DESY 97-203 \\ HUB-EP-97/72 \vspace*{18cm}

\section{METHOD}

The renormalization method we study is a $MOM$-scheme using the lattice as
regulator with the lattice spacing $a$ as cut-off.  
Lattice operators ${\cal O}(a)$ and quark fields $q(a)$ are 
renormalized being then only a
function of the scale parameter $\mu$, via:
\begin{eqnarray}
   {\cal O}^R(\mu) &=& Z_{\cal O}((a\mu)^2, g(a)) {\cal O}(a),
                                            \nonumber \\
   q^R(\mu) &=& Z_q^{1/2}((a\mu)^2, g(a)) q(a) \nonumber
\end{eqnarray}
($g(a)$ is the bare lattice coupling.) 

To determine the
renormalization constants $Z_{\cal O}$ one 
imposes conditions on amputated forward
quark vertex functions in momentum space
with external four-momenta $p$:
\begin{eqnarray} 
  \langle q(p) | {\cal O}(\mu) | q(p)  \rangle^R_{amp} 
 &\!\!\!\!\!\equiv& \!\!\!\!
  \langle q(p) | {\cal O}(a) | q(p)  \rangle^{tree}_{amp} 
 \left. \!\right|_{p^2=\mu^2}  \nonumber 
\end{eqnarray}
With a projector method, where
$Tr[\Gamma_{\cal O}\!\times ...]$ is taken,
$Tr$ being a colour$\times$spin trace, 
these prescriptions give:
\begin{eqnarray} \lefteqn{ \!\!\!\!\!\!\!\!\!\!\!
  Z_{\cal O} Z_q^{-1} Tr[ \big. \Gamma_{\!\!\cal O}
  \langle q(p) | {\cal O}(a) | q(p)  \rangle_{amp} \big. ]
  } \nonumber \\
  \qquad  & = & 
   Tr[ \big. \Gamma_{\!\!\cal O}
  \langle q(p) | {\cal O}(a) | q(p)  \rangle^{tree}_{amp} 
  ] \big|_{p^2=\mu^2}          
\end{eqnarray}
As an optimal choice for the projector, we use, written in a general 
form: 
\[ \Gamma_{\cal O}\!\!\sim\!\langle q(p)|
{\cal O}(a)|q(p)\rangle^{tree}_{amp} \] 

There are two definitions for $Z_q$:
Usually one projects onto the energy-momentum part 
$\sin (p_\mu a)\gamma_\mu/a$ (continuum form: $\pslash$) of
the inverse propagator $S^{-1}$ and gets: \\  

\mbox{ \hspace{-0.59cm}
  \begin{math} Z_q\!=\!Tr \Big[ \Big.
    \frac{-i \sum_\lambda \gamma_\lambda \sin(p_\lambda a) }
    {(2 \kappa) 12 \sum_\lambda \sin^2(p_\lambda a)  } S^{-1}(p_\lambda a)
   \Big. \Big]_{p^2=\mu^2}
  \end{math}}    \\

{ \hspace{-0.33cm}(The normalization is chosen to give $Z_q=1$ in the 
   free case.)}
Another definition for $Z_q$ can be derived from eq. (1), setting 
${\cal O}$ equal to the conserved vector current $J_\mu$ and
using \mbox{$Z_{J_\mu}=1$}.
But this is only compatible to the first determination of  $Z_q$,
if the component of $J_\mu$ transverse to $p$ is taken.
(This can be proven by a Ward Identity 
\mbox{${J_\mu}(p,p) = -i \partial / \partial p_\mu S^{-1}(p)$} 
assuming 
\mbox{$S^{-1}(p)= i Z_q(p^2) \pslash + B(p^2) 1$}, which is true
in the continuum and 
approximately true on the lattice for not too large
$p^2a^2$.)
%

\section{NUMERICAL IMPLEMENTATION}
In a first step
Monte Carlo gluon field configurations $U$ have to be generated
and numerically fixed to the Landau gauge \cite{4}. The lattice matrix
elements in eq. (1), non-amputated, can then be calculated in 
those background fields:
\begin{eqnarray} \lefteqn{ \!\!\!\!\!\!\!\!\!\!\!
  {\langle q(p) | {\cal O}(a) | q(p)  \rangle}
  } \nonumber \\
\qquad  &\sim& 
   \sum_{i,j_1,j_2,k}
  {\langle S_{i,j_1} O_{j_1,j_2} S_{j_2,k} \rangle}_U 
   e^{i p (x_k - x_i)} 
\end{eqnarray}
($i,j_1$,... are space-time points on the lattice.) 

The propagators
$S_{j_2,p}=\sum_k S_{j_2,k}e^{i p x_k}$ can be computed 
from a lattice Dirac equation with a momentum source, 
$M$ being the fermion matrix:
\begin{eqnarray}
 \sum_k M_{i,k} S_{k,p} & = & e^{i p x_i} 
               \qquad \mbox{(momentum source)} \nonumber
\end{eqnarray}
So here the number of matrix inversions
is proportional to the number of momenta. But 
everything else is then there:  
    $S_{p,j_1}  = \gamma_5 S_{j_1,p}^{\dagger} \gamma_5 $ and
    the quark propagator $S(p) =  {\langle S_{p,p}\rangle}_U$ with 
    $S_{p,p}=\sum_k S_{p,k}e^{i p x_k}$ for an amputation and
$Z_q$.

The momentum source method automatically performs all the site
sums in eq. (2). Another possibility, instead of summing over $j_2$
(and $j_1$), is to choose a particular location for the operator,
for example setting $j_2=0$. Translational invariance tells us that
this will give the same expectation value after averaging over all
configurations. For this method we need to solve the Dirac equation 
with a point source at $j_1=j_2$ and (for extended operators where 
$j_1 \neq j_2$) for a small number of sources in the region around 
$j_2$: 
\begin{eqnarray}
 \sum_k M_{i,k} S_{k,j_1} & = & \delta_{i,j_1}
               \qquad \mbox{(point source)} \nonumber
\end{eqnarray}
($S_{k,j_1}$ has simply to be fourier 
transformed to $S_{p,j_1}$ then 
and we have $S(p) \sim \sum_{j_1} {\langle S_{p,j_1}\rangle}_U$.)

For local operators and operators with a small number of derivatives
the point source method would need fewer inversions, but we see from
Fig.1. that relying on translational invariance to carry out the $j$
sums leads to much larger error bars.


\begin{figure}[h]
   \vspace*{-1.10cm}   
   \hspace*{-0.50cm}
   \epsfxsize=8.00cm \epsfbox{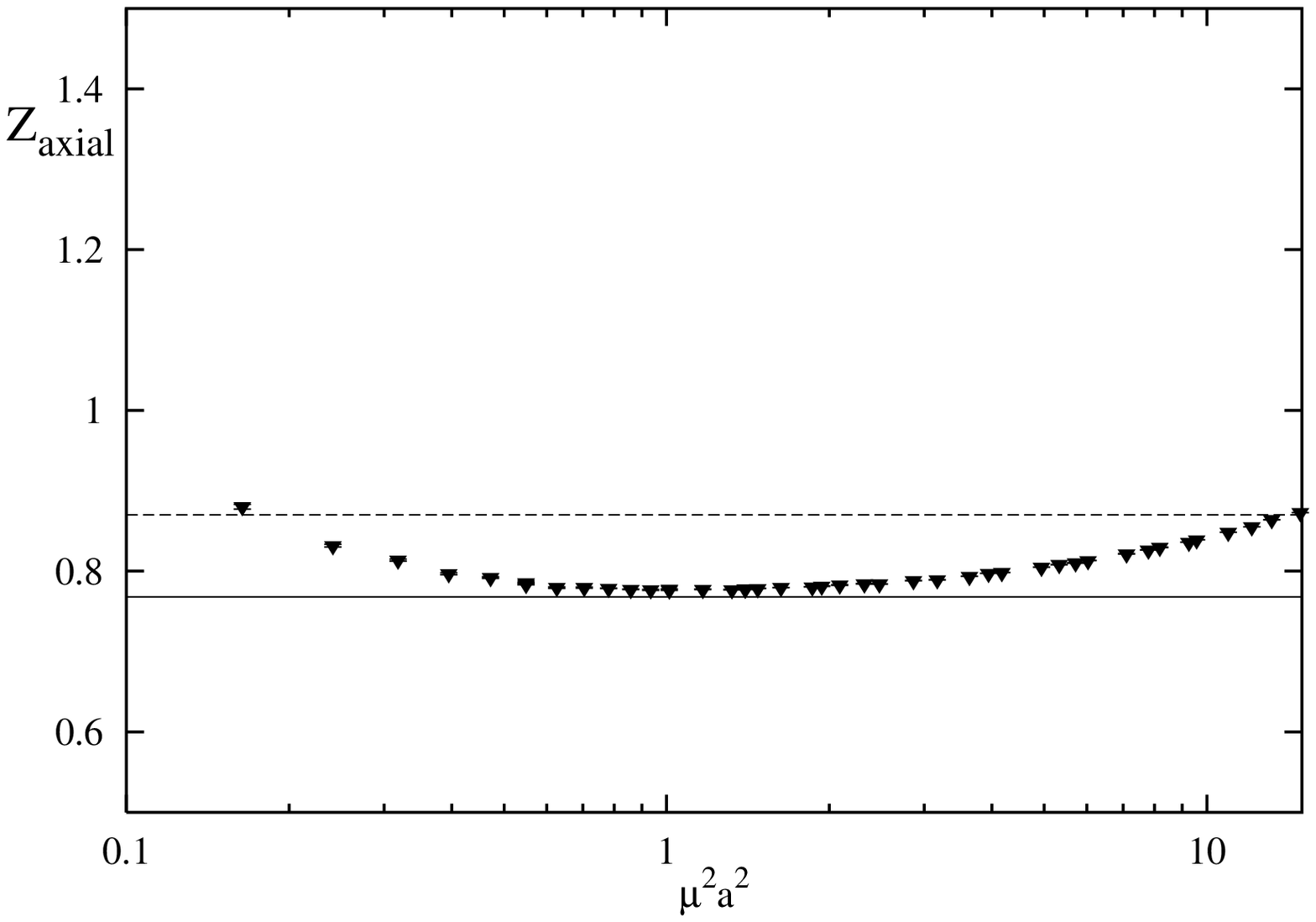}  
   \vspace*{-1.25cm}
   \vspace*{-0.65cm}      
   \label{fig1}
\end{figure}

\begin{figure}[h]
   \vspace*{-0.68cm}   
   \hspace*{-0.50cm}
   \epsfxsize=8.00cm \epsfbox{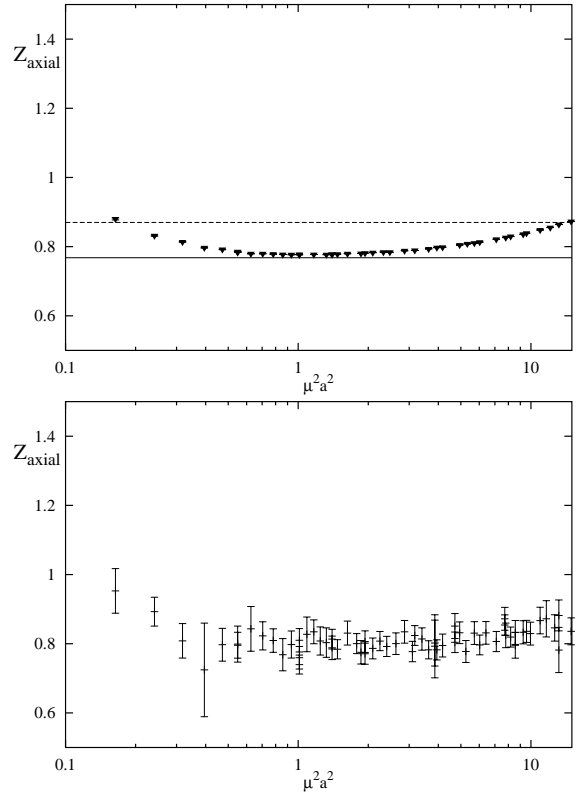}  
   \vspace*{-1.10cm}   
   \caption{\footnotesize $Z_{axial}$
         with a momentum source method (top figure; 20 config.) and a 
         point source method (120 config.)}
   \vspace*{-1.00cm}        
   \label{fig2}
\end{figure}

\section{RESULTS (WILSON FERMIONS)}

Our calculations are performed on $16^332$ lattices at $\beta\!=\!6.0$ for 
hopping parameters $\kappa\!=\!0.1515$, $0.1530$, $0.1550$.
If the method works for current lattices, one expects to find a
high enough $\mu^2a^2$ region where the $Z$ factors agree with
\mbox{lattice per-}
%
\newpage

\begin{figure}[h]
   \vspace*{-0.25cm}   
   \hspace*{-0.50cm}
   \epsfxsize=8.00cm \epsfbox{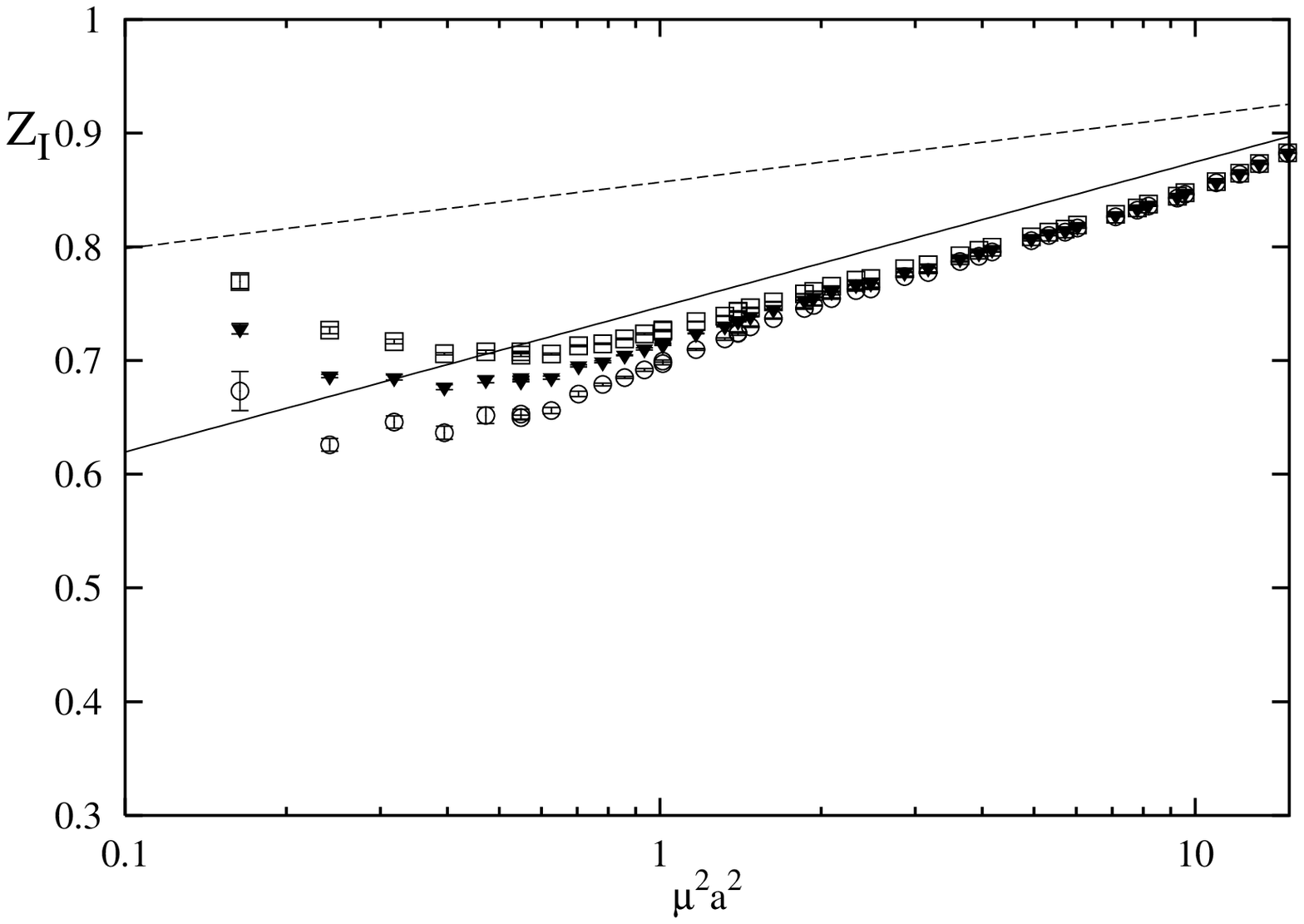}  
   \vspace*{-0.90cm}       
   \vspace*{-0.75cm}     
   \label{fig3}
\end{figure}

\begin{figure}[h]
   \vspace*{-1.00cm}   
   \hspace*{-0.50cm}
   \epsfxsize=8.00cm \epsfbox{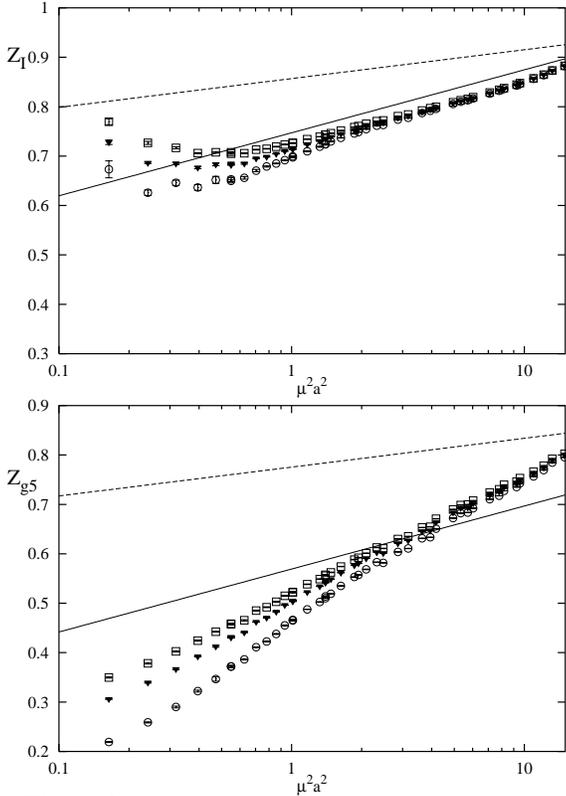}  
   \vspace*{-1.15cm}
   \caption{\footnotesize $Z$ of the scalar and pseudo-scalar operator
            for $\kappa=0.1515$, $0.1530$, $0.1550$  
            with lower points belonging to highest $\kappa$  
            (lightest quark mass) are shown.   }
   \vspace*{-0.75cm}
   \label{fig4}
\end{figure}

\mbox{\hspace{-0.44cm} turbation theory}, but are still not destroyed
by discretisation effects. However the available 
one-loop calculations
probably need to be improved for $g^2=6/\beta=1$.
At low $\mu^2a^2$ non-perturbative
effects and higher order perturbation theory may have a 
big influence as can 
be seen looking at the pseudo-scalar $Z_{g5}$ (Fig.2):
results for $3$ quark masses indicate a
non-perturbative pion pole contribution in the chiral limit. 
The $Z$ factors for the operators of the moments of the 
structure functions, e.g. 
$Z_{\langle x \rangle}$ (Fig.3), seem to show also big
non-perturbative or higher order perturbation theory contributions 
at low $\mu^2a^2$. 
O(a) discretization errors are in general expected at 
higher $\mu^2a^2$. However, for the local vector current (Fig.3)
such errors seem to be there at all scales: it is not constant as
expected from its continuum behaviour.
In the O(a) improved theory \cite{2} it is.
 
In a 
comparison with
one-loop lattice perturbation theory 
(dashed lines in figures) \cite{5}: \\
$ Z_{\cal O}
      = 1 - g^2 /(16\pi^2) C_F
                         ( \gamma_{\cal O} \ln(a\mu)
                           + B_{\cal O}
                         ) $,            \\
$C_F=4/3$ and $\gamma_{\cal O}$ being the anomalous dimension,
and with tadpole improved theory (solid lines in figures)
\cite{6}   
we find that the improved theory fits better to the
data. For the axial current (Fig.1) and the scalar operator (Fig.2)
there is a good agreement.
We extract $Z_{axial}=0.78$.

\begin{figure}[h]
   \vspace*{-1.05cm}   
   \hspace*{-0.50cm}
   \epsfxsize=8.00cm \epsfbox{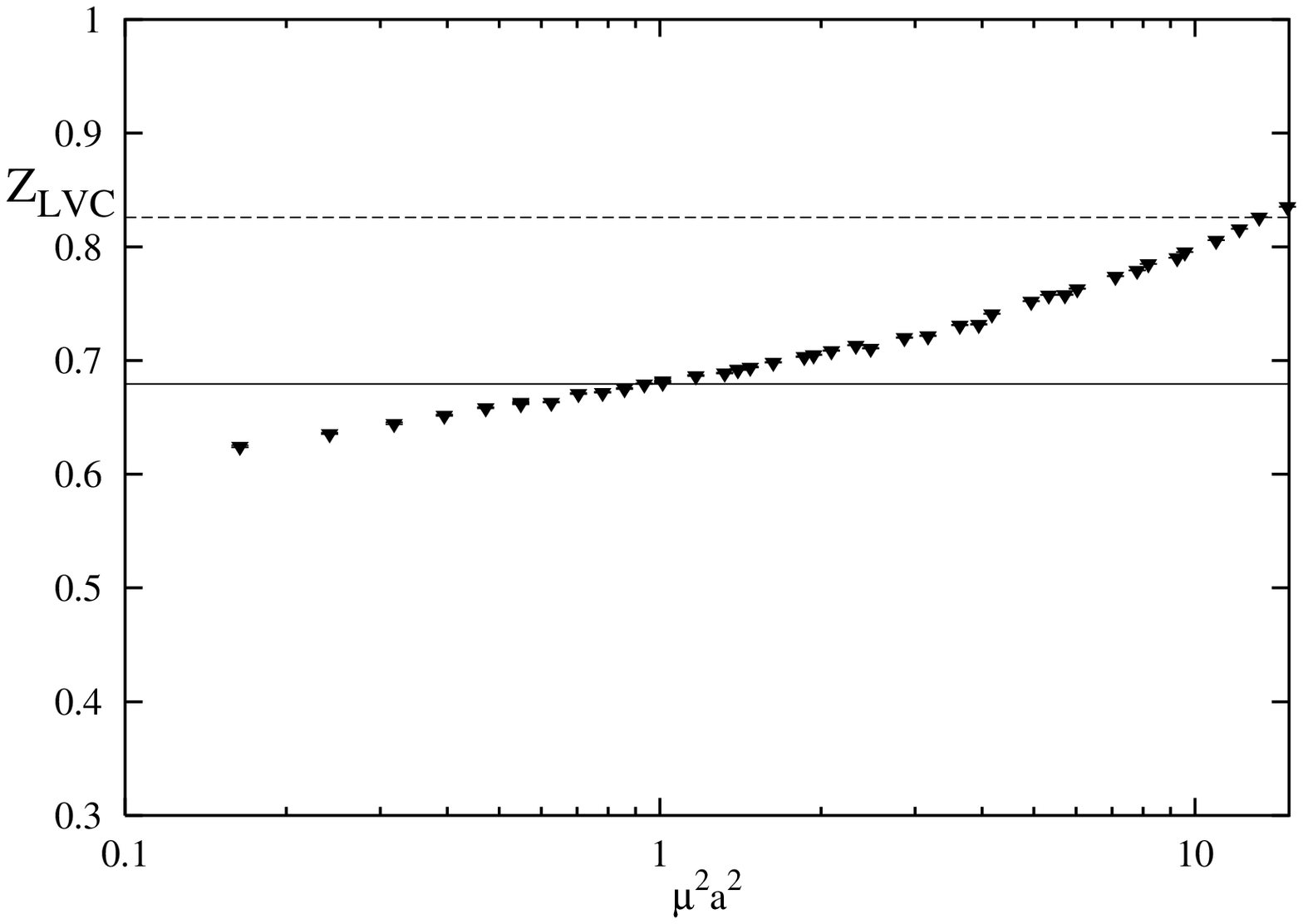}  
   \vspace*{-1.25cm}
   \vspace*{-0.75cm}
   \label{fig5}
\end{figure}
\begin{figure}[h]
   \vspace*{-0.70cm}   
   \hspace*{-0.50cm}
   \epsfxsize=8.0cm \epsfbox{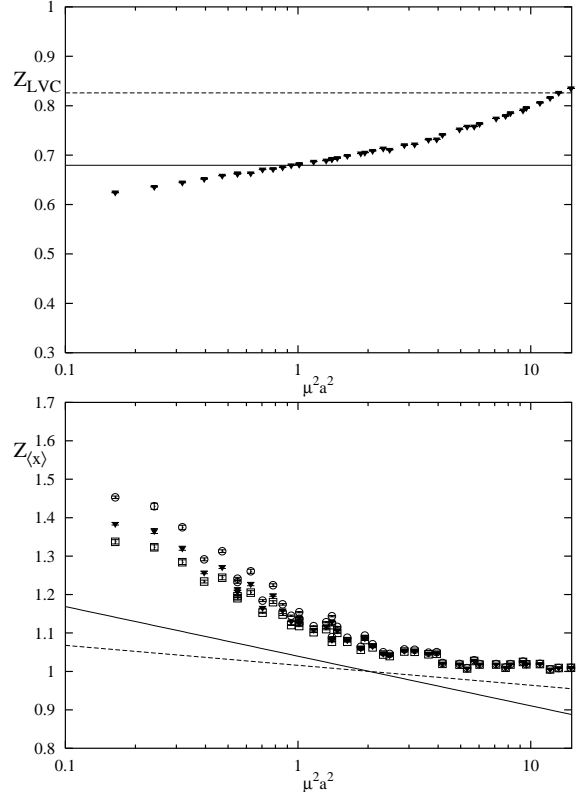}  
   \vspace*{-1.15cm}     
  \caption{\footnotesize $Z$ of the local vector 
current LVC ($\kappa\!=\!0.1530$) and  
$Z_{\langle x\rangle}$ 
            ($3$ $\kappa$,  
            lower points belong to lowest $\kappa$) are shown. }
   \vspace*{-0.80cm}    
   \label{fig6}
\end{figure}

\end{document}